# Spectrum averaged cross section measurements of lutetium using standard $^{252}$Cf neutron source

Martin Schulc[a], Michal Košťál[a], Tomáš Czakoj[a], Evžen Novák[a], Jan Šimon[a], Nella Hynková[a] and Roberto Capote[b]

[a] Research Centre Rez Ltd, 250 68 Husinec-Řež 130, Czech Republic
[b] Nuclear Data Section, International Atomic Energy Agency, A-1400 Vienna, Austria

Email: Martin.Schulc@cvrez.cz

Abstract
The spectrum averaged cross section (SACS) in standard neutron field is a preferable tool for cross section validation. There are very few measurements involving lutetium neutron cross sections. The presented work uses only neutron standard, i.e., $^{252}$Cf(sf) neutron field, for validation of lutetium threshold cross sections. SACS were inferred from gamma spectrometry derived reaction rates. The SACS which were derived include $^{175}$Lu(n,2n)$^{174}$Lu, $^{175}$Lu(n,3n)$^{173}$Lu, $^{175}$Lu(n,p)$^{175}$Yb, and $^{176}$Lu(n,n′)$^{176m1}$Lu reactions. All these reactions SACS were measured for the first time. MCNP6 calculations using JEFF-3.3 or ENDF/B-VIII.0 libraries for lutetium cross sections were compared with experimental data. The agreement was found very poor for all reactions under study. Thus there is a need for their improvement. The presented data can be also used for validation of the various theoretical models.

## 1 Introduction

Lutetium is a rare last element in the lanthanide series. Natural lutetium consists of two isotopes, $^{175}$Lu (97.401%) and $^{176}$Lu (2.599%). Isotope $^{176}$Lu is radioactive, having a half-life of 3.78E10 years. This isotope and its ratio to its decay product $^{176}$Hf is used to geochronical dating of minerals and rocks. $^{175}$Lu, having a relatively high cross-section for (n,xn) reactions, is present in the NEA Nuclear Data High Priority Request List (Plompen et al., 2007) with a possibility to be used for dosimetry purposes. Despite this fact, the differential experimental data on lutetium are scarce; furthermore, the integral data are entirely missing. This work deals with the measurement of lutetium $^{252}$Cf(sf) spectrum averaged cross sections (SACS), which is an integral quantity useful for validation. The presented experimental data for lutetium will help to complement incomplete nuclear data. Due to the well known $^{252}$Cf(sf) neutron spectrum, beimg primary neutron standard, these data also can help to understand the nuclear reaction mechanisms and, in general, neutron-nuclear interactions. Also, it can be helpful for testing different theoretical models; for example statistical model (Dzysiuk et al., 2010). The neutron standard $^{252}$Cf(sf) was used for validation of lutetium threshold cross sections. The investigated SACS include $^{175}$Lu(n,2n)$^{174}$Lu, $^{175}$Lu(n,3n)$^{173}$Lu, $^{175}$Lu(n,p)$^{175}$Yb, and $^{176}$Lu(n,n′)$^{176m1}$Lu reactions. The methodology of SACS evaluation and $^{252}$Cf source handling was successfully validated in previously published papers such as (Schulc et al. 2018a, Schulc et al. 2019a, Schulc et al. 2019b, and Schulc et al. 2020).

## 2 Description of the experimental setup and evaluations

The used neutron source, $^{252}$Cf, had an average total emission of 1.93E8 n/s during lutetium irradiation. The source emission was derived from the data in the Certificate of Calibration received from the National Physical Laboratory, United Kingdom. The thin nickel foils were attached in front of and behind the lutetium foil as a neutron flux monitor due to the complicated geometry. The agreement between nickel inferred emission and calculated emission was found. The lutetium activation foil (2.5 cm × 2.5 cm, 0.1 cm thick) was attached to the surface of the irradiation tube to achieve maximal neutron flux, see Figure 1 for the experimental setup. The centre of the source in the tube was leveled with a centre of the foil. Irradiation of the lutetium took almost 99 days continuously. After the irradiation, the foil was subsequently measured on the upper cap of the high-purity germanium detector. The efficiency curve of the detector was determined calculationally using validated MCNP6.2 (Goorley et al., 2012) model, see (Kostal et al. 2017). The gamma spectrum acquisition took almost 4 days. Table 1 summarizes the parameters of the irradiation and gamma spectrometry. Table 2 displays measured and evaluated gamma lines of the activation products, their emission probabilities, and their half-lives.

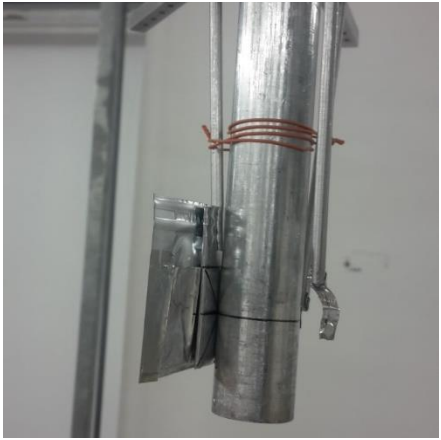

Figure 1: Lutetium foil irradiation setup.

*Table 1: Parameters of irradiation and following HPGe measurement.*

| Reaction | Irradiation time | Cooling time | Measurement time |
|---|---|---|---|
| All reactions | 98.91 days | 10 minutes | 3.84 days |

*Table 2: Parameters of the investigated neutron-induced threshold reactions.*

| Reaction | Half-life | Gamma Energy [MeV] | Gamma emission probability |
|---|---|---|---|
| $^{175}$Lu(n,2n)$^{174g}$Lu | 3.31 years | 1.241847 | 5.14% |
| $^{175}$Lu(n,2n)$^{174m}$Lu | 142 days | 0.067058 | 7.25% |
| $^{175}$Lu(n,3n)$^{173}$Lu | 1.37 years | 0.272105 | 21.2% |
| $^{175}$Lu(n,p)$^{175}$Yb | 4.185 days | 0.396329 | 13.2% |
| $^{176}$Lu(n,n′)$^{176m1}$Lu | 3.664 hours | 0.088361 | 8.9% |

The experimental reaction rate $q$ was calculated from the measured gamma line net peak area using formula

$$q = \frac{C(T_m)\lambda T_m}{\eta \varepsilon N k T_l} \frac{1}{e^{-\lambda \Delta T}} \frac{1}{1-e^{-\lambda T_m}} \frac{1}{1-e^{-\lambda T_{irr}}}, \qquad (1)$$

where: $q$ is the experimental reaction rate per atom per second, $N$ is the number of target isotope nuclei, $\eta$ is the detector efficiency, $\varepsilon$ is gamma branching ratio, $\lambda$ is the decay constant, $k$ characterizes the abundance of isotope of interest in the target and its purity, $\Delta T$ is the time between the end of irradiation and start of HPGe measurement, $C(T_m)$ is the measured number of counts, $T_m$ is the real time of measurement by HPGe, $T_l$ is the live time of measurement by HPGe (it is time of measurement corrected to the dead time of the detector), and $T_{irr}$ is the time of irradiation. The coincidence summing was estimated to be sufficiently low and, therefore, it was neglected.

The SACS is derived from reaction rate $q$ by correction factor $C$ which considers the spectral shift effect, flux loss and self-shielding together. The correction is computed by means of MCNP6 as a ratio between the SACS in the real set-up and the SACS in the same set-up consisting of void cells. The $^{252}$Cf SACS is derived via Equation 2:

$$\bar{\sigma} = \frac{\int_E \sigma(E)\,\varphi(E)\,dE}{\int_E \varphi(E)\,dE} \times C, \qquad (2)$$

where $C$ denotes the correction factor, $\varphi(E)$ is the calculated neutron spectrum, $\sigma(E)$ is the cross section and $\bar{\sigma}$ is the spectral averaged cross section. The numerator corresponds to the measured reaction rate according to the Equation 1.

The uncertainty analysis includes non-negligible uncertainties: uncertainty in the position of the sample, emission and the position of the $^{252}$Cf source, statistical uncertainty of the net peak area, and the calculated germanium detector efficiency uncertainty, details can be found in (Schulc et al. 2019b).

MCNP6.2 software was used for all calculations. The base transport libraries were taken from ENDF/B-VII.1 (Chadwick et al. 2011). Lutetium cross-section was taken either ENDF/B-VIII.0 (Brown et al. 2018) or JEFF-3.3 (Plompen et al., 2020). The input $^{252}$Cf(sf) neutron spectrum was taken from (Mannhart, 2008). The calculations were performed using all dimensions, densities, and materials.

## 3  Results

Figure 2 shows the comparison of $^{175}$Lu(n,2n)$^{174}$Lu reaction cross-section with the experimental data available in the EXFOR database. Figure 3 displays the energy distribution of the reaction rate for $^{175}$Lu(n,2n)$^{174}$Lu in JEFF-3.3 and ENDF/B-VIII.0 libraries. The distribution in both libraries is similar. The cross-section in both libraries is also similar in the region under 14 MeV, i.e. the most sensible energy region. The calculation in both libraries gives a similar result and overestimates the experiment around 20 %. The $^{175}$Lu(n,2n)$^{174g+m}$Lu reaction SACS in $^{252}$Cf was derived as 7.02 ± 0.26 mb. This result was obtained as the sum of SACS of $^{175}$Lu(n,2n)$^{174g}$Lu and $^{175}$Lu(n,2n)$^{174m}$Lu because both channels were measured (see Table 2).

In the case of the $^{175}$Lu(n,3n)$^{173}$Lu reaction, Figure 4 shows the reaction cross-section. JEFF-3.3 library differs significantly from the ENDF/B-VIII.0 library. JEFF-3.3 library follows EXFOR data, unlike ENDF/B-VIII.0. Figure 5 shows the reaction rate contribution of $^{175}$Lu(n,3n)$^{173}$Lu reaction.

The distributions differ; however, a maximal response is in the same energy bin for both libraries (17-18 MeV). The agreement with the experiment is not satisfactory, the JEFF-3.3 library underestimates the experiment around 75 % and the ENDF/B-VIII.0 library around 45 %. The SACS for the $^{175}$Lu(n,3n)$^{173}$Lu reaction in the $^{252}$Cf spectrum was derived as 0.131 ± 0.008 mb.

Concerning $^{175}$Lu(n,p)$^{175}$Yb reaction, Figure 6 shows its cross-section. Evaluations in JEFF-3.3 and ENDF/B-VIII.0 are significantly different. The same situation is also for the reaction rate distribution depicted in Figure 7. Note that data in the EXFOR database are contradictory. It is apparent that the JEFF-3.3 and ENDF/B-VIII.0 are based on different EXFOR experimental data. The agreement of the present experiment with calculation is very bad, -83 % in the case of JEFF-3.3 library and approximately 725 % for ENDF/B-VIII.0 library. $^{175}$Lu(n,p)$^{175}$Yb reaction SACS value is the lowest from all studied reactions, i.e. 6.81E-02 ± 0.23E-02 mb.

The last explored reaction was $^{176}$Lu(n,n´)$^{176m1}$Lu reaction. In this case, the unexcited $^{176}$Lu isotope has gamma line of very similar energy as its first excited state (unexcited $^{176}$Lu 0.08834 MeV and excited $^{176}$Lu 0.088361 MeV). The natural activity was subtracted from measured activity in the SACS evaluation. The cross-section leading to first excited state was not found in any library, neither any data in EXFOR. Nevertheless, still some results can be inferred from inelastic cross section (leading to all excited states). Comparison of measured $^{176}$Lu(n,n´)$^{176m1}$Lu with calculated $^{176}$Lu(n,n´)$^{176}$Lu gives C/E-1 ratio around -70 %. This fact means that inelastic cross-section is necessarily inconsistent with the experiment. Figure 8 compares $^{176}$Lu(n,inl)$^{176}$Lu reaction in available libraries. $^{176}$Lu(n,n´)$^{176m1}$Lu reaction SACS value was estimated as 7.10E03 ± 0.22E3 mb.

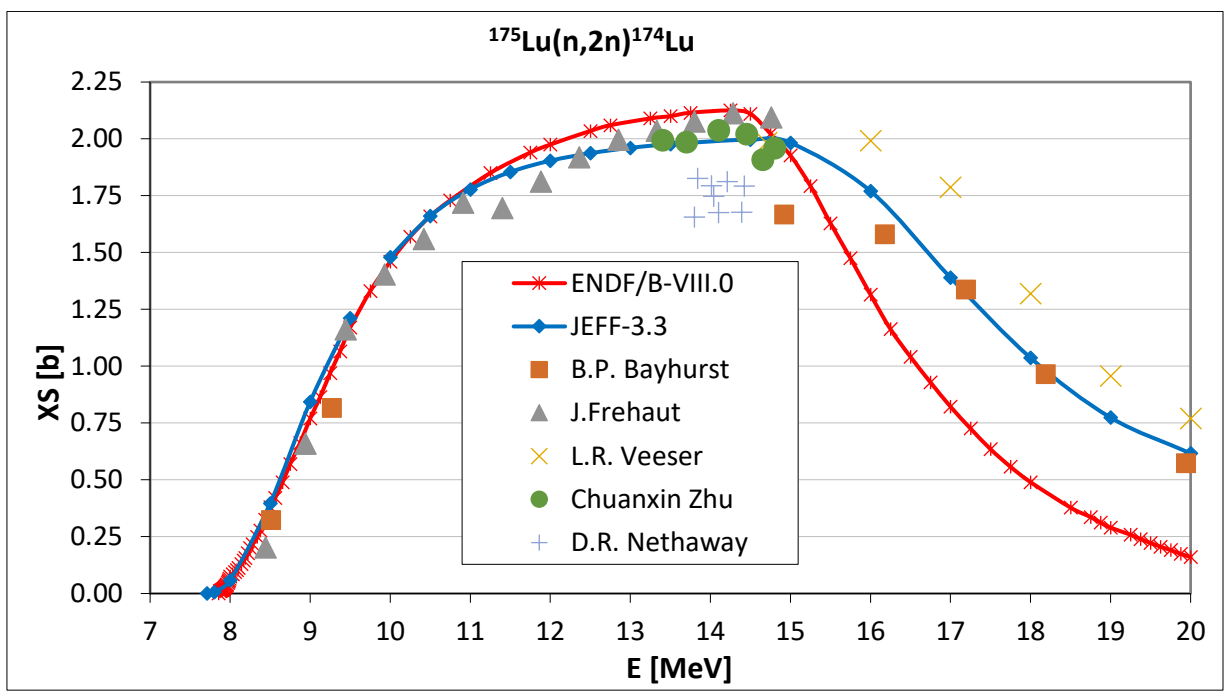

Figure 2: Comparison of $^{175}$Lu(n,2n)$^{174}$Lu reaction cross-section with the EXFOR data, JEFF-3.3 and ENDF/B-VIII.0 libraries.

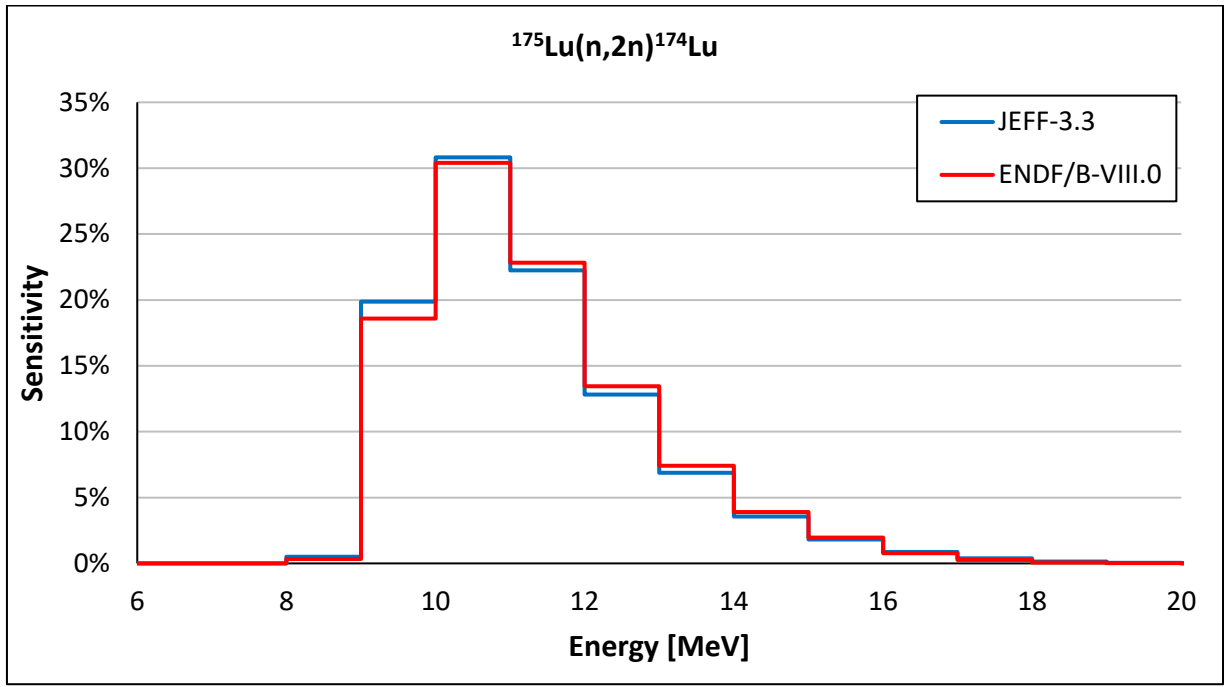

Figure 3: Energy distribution of the reaction rate for $^{175}$Lu(n,2n)$^{174}$Lu in JEFF-3.3 and ENDF/B-VIII.0 libraries.

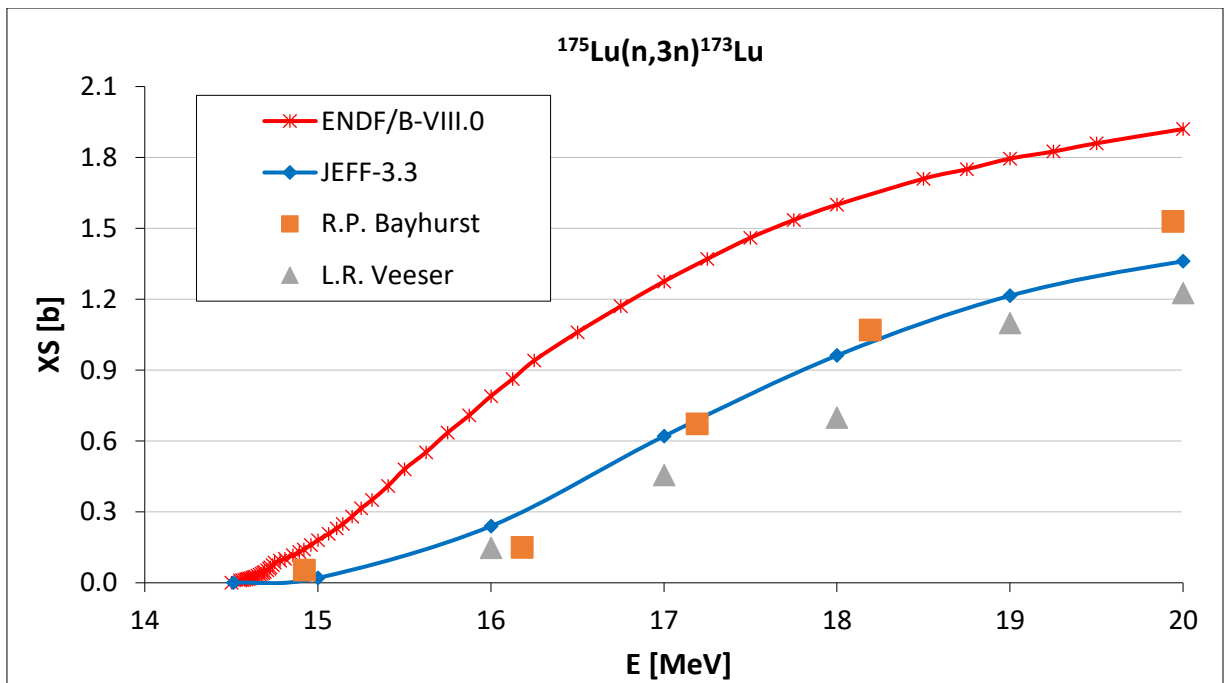

Figure 4: Comparison of $^{175}$Lu(n,3n)$^{173}$Lu reaction cross-section with the EXFOR data, JEFF-3.3 and ENDF/B-VIII.0 libraries.

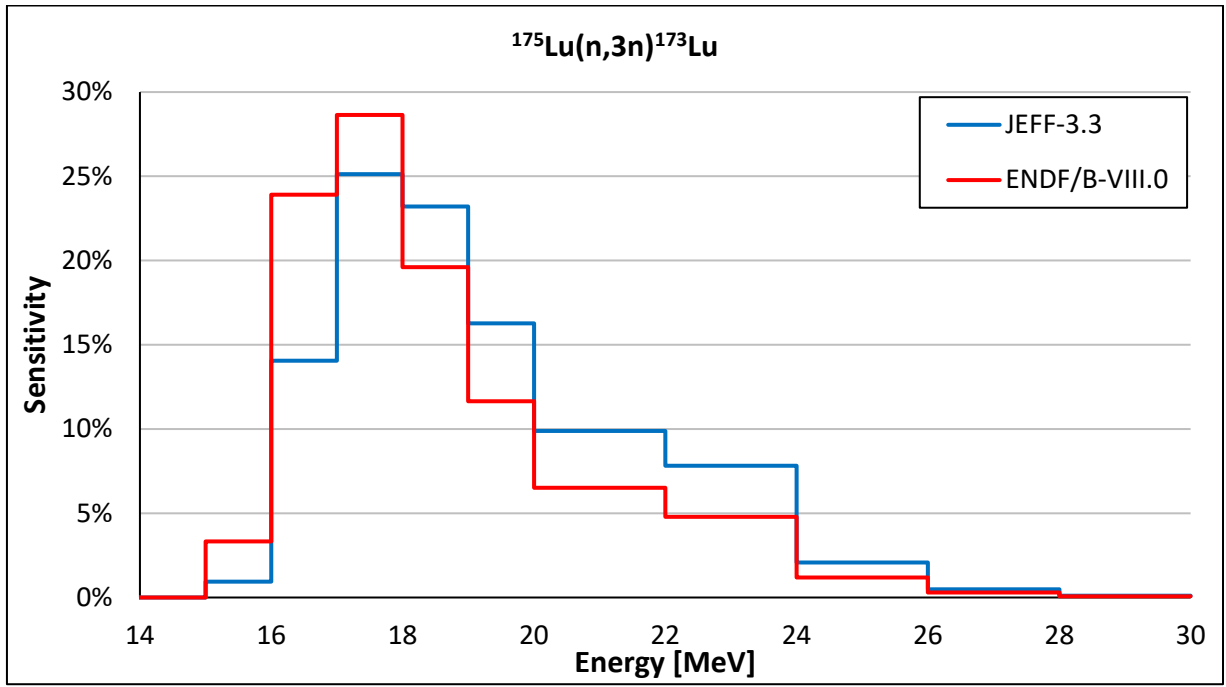

Figure 5: Energy distribution of the reaction rate for $^{175}$Lu(n,3n)$^{173}$Lu in JEFF-3.3 and ENDF/B-VIII.0 libraries.

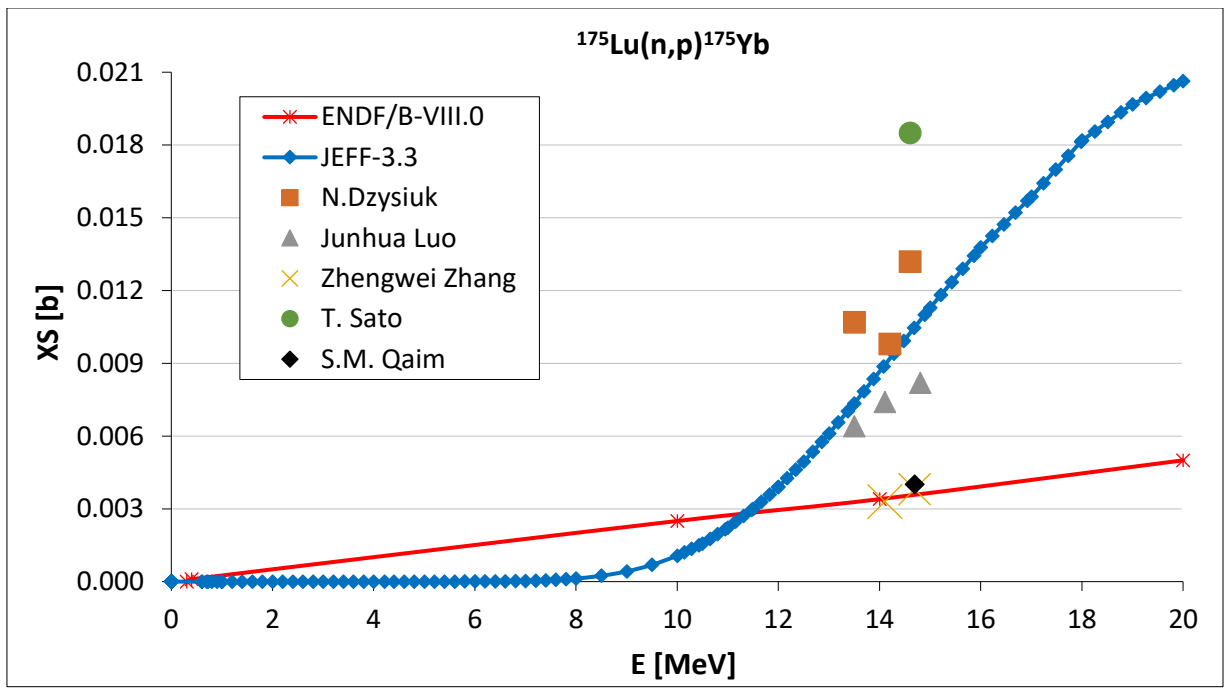

Figure 6: Comparison of $^{175}$Lu(n,p)$^{175}$Yb reaction cross-section with the EXFOR data, JEFF-3.3 and ENDF/B-VIII.0 libraries.

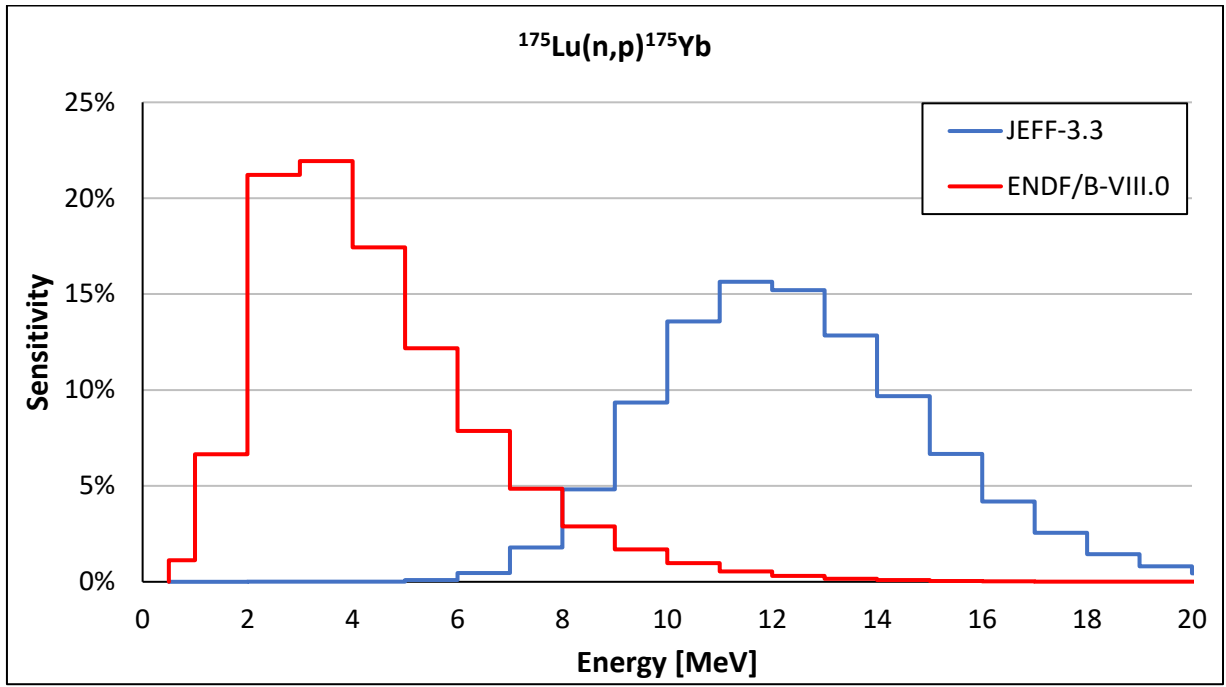

Figure 7: Energy distribution of the reaction rate for $^{175}$Lu(n,p)$^{175}$Yb in JEFF-3.3 and ENDF/B-VIII.0 libraries.

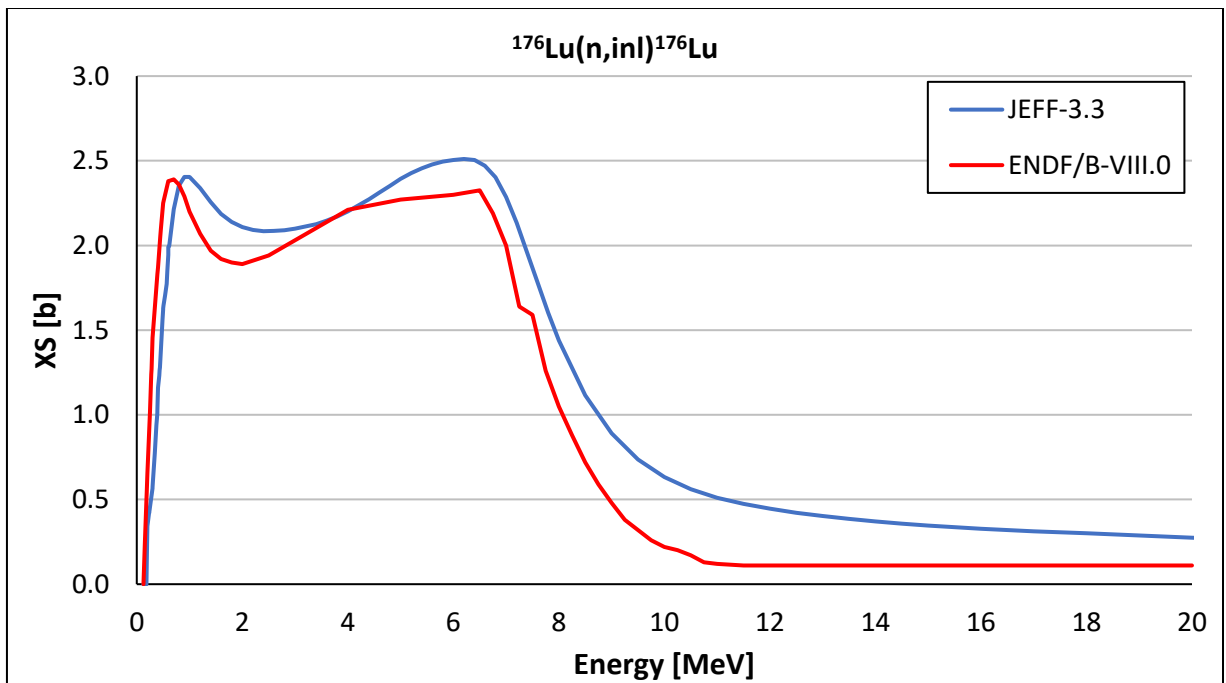

Figure 8: Comparison of $^{176}$Lu(n,inl)$^{176}$Lu reaction cross-section in JEFF-3.3 and ENDF/B-VIII.0 libraries.

## 4  Conclusions

Table 3 summarizes experimental and calculated reaction rates for all reactions. The performed experiment reveals unsatisfactory agreement with any libraries cross-section for lutetium neutron induced threshold reactions. These cross-sections need further improvement and more experiments dealing with lutetium to be performed. All SACS were measured for the first time.

*Table 3: Calculation and C/E-1 comparison for neutron induced threshold reactions in lutetium.*

| $^{175}$Lu(n,2n)$^{174}$Lu | Q [s$^{-1}$ atom$^{-1}$ neutron$^{-1}$] | C/E-1 | Uncertainty |
|---|---|---|---|
| EXPERIMENT | 2.00E-28 | | 3.7% |
| JEFF-3.3 | 2.45E-28 | 22.26% | |
| ENDF/B-VIII.0 | 2.39E-28 | 19.10% | |
| $^{175}$Lu(n,3n)$^{173}$Lu | | | |
| EXPERIMENT | 3.74E-30 | | 5.8% |
| JEFF-3.3 | 9.31E-31 | -75.14% | |
| ENDF/B-VIII.0 | 2.05E-30 | -45.35% | |
| $^{175}$Lu(n,p)$^{175}$Yb | | | |
| EXPERIMENT | 1.94E-30 | | 3.4% |
| JEFF-3.3 | 3.30E-31 | -83.04% | |
| ENDF/B-VIII.0 | 1.60E-29 | 725.01% | |
| $^{176}$Lu(n,n´)$^{176m1}$Lu | | | |
| EXPERIMENT | 2.27E-25 | | 3.1% |

Acknowledgement

Presented results were obtained with the use of infrastructure Reactors LVR-15 and LR-0, which is financially supported by the Ministry of Education, Youth and Sports – project LM2018120 and the SANDA project funded under H2020-EURATOM-1.1 contract 847552.